\documentclass[iop,apj]{emulateapj}
\usepackage{apjfonts}
\usepackage{graphicx,float,wrapfig}
\usepackage{multirow, threeparttable, color}
\usepackage{hyperref}
\hypersetup{
    colorlinks,
    citecolor=blue,
    filecolor=blue,
    linkcolor=blue,
    urlcolor=blue
}

\makeatletter
\newcommand*{\rom}[1]{\expandafter\@slowromancap\romannumeral #1@}
\makeatother 

\newcommand{\fesc}{\ifmmode{f_{\rm esc}}\else{$f_{\rm esc}$}\fi}
\newcommand{\fescs}{\ifmmode{f_{\rm esc}^\star}\else{$f_{\rm esc}^\star$}\fi}
\newcommand{\kms}{\ifmmode{{\;\rm km~s^{-1}}}\else{km~s$^{-1}$}\fi}
\newcommand{\fgas}{\ifmmode{{f_{\rm gas}}}\else{$f_{\rm gas}$}\fi}
\newcommand{\cubecm}{\ifmmode{{\rm cm^{-3}}}\else{cm$^{-3}$}\fi}
\newcommand{\ztwo}{\ifmmode{{\rm [Z_2/H]}}\else{[Z$_2$/H]}\fi}
\newcommand{\zthree}{\ifmmode{{\rm [Z_3/H]}}\else{[Z$_3$/H]}\fi}
\newcommand{\lsim}{\lower0.3em\hbox{$\,\buildrel <\over\sim\,$}}
\newcommand{\gsim}{\lower0.3em\hbox{$\,\buildrel >\over\sim\,$}}

\newcommand{\sfr}{\ifmmode{\textrm{M}_\odot \,\textrm{yr}^{-1} \,\textrm{Mpc}^{-3}}\else{M$_\odot$ yr$^{-1}$ Mpc$^{-3}$}\fi}
\newcommand{\hsfr}{\ifmmode{\;\textrm{M}_\odot\, \textrm{yr}^{-1}}\else{M$_\odot$ yr$^{-1}$}\fi}

\newcommand{\eavg}{\ifmmode{\langle E_\gamma \rangle}\else{$\langle E_\gamma \rangle$}\fi}

\newcommand{\enzo}{{\it Enzo}}
\newcommand{\yt}{{\it yt}}

\newcommand{\Ms}{\ifmmode{M_\odot}\else{$M_\odot$}\fi}
\newcommand{\vrms}{\ifmmode{v_{\rm rms}}\else{$v_{\rm rms}$}\fi}

\newcommand{\tvir}{\ifmmode{T_{\rm{vir}}}\else{$T_{\rm{vir}}$}\fi}
\newcommand{\mvir}{\ifmmode{M_{\rm{vir}}}\else{$M_{\rm{vir}}$}\fi}
\newcommand{\rvir}{\ifmmode{r_{\rm{vir}}}\else{$r_{\rm{vir}}$}\fi}

\newcommand{\jj}{\ifmmode{J_{21}}\else{$J_{21}$}\fi}
\newcommand{\flw}{\ifmmode{F_{LW}}\else{$F_{LW}$}\fi}
\newcommand{\kph}{\ifmmode{k_{\rm ph}}\else{$k_{\rm ph}$}\fi}

\newcommand{\zsun}{\ifmmode{\rm\,Z_\odot}\else{$\rm\,Z_\odot$}\fi}

\newcommand{\nhi}{\ifmmode{N_{\rm HI}}\else{$N_{\rm HI}$}\fi}

\begin{document}

\shorttitle{Songlines from Black Holes}
\shortauthors{Aykutalp et al.}
\journalinfo{Submitted to the Astrophysical Journal}
\submitted{Submitted 2014 August 29;  accepted 2014 October 7}

\title{Songlines from Direct Collapse Seed Black Holes: Effects of X-rays on Black Hole Growth and Stellar Populations}
\author{Aycin Aykutalp\altaffilmark{1,2},
  John H. Wise\altaffilmark{1}, 
  Marco Spaans\altaffilmark{3}, 
  and Rowin Meijerink\altaffilmark{4}}

\affil{$^{1}${Center for Relativistic Astrophysics, School of
  Physics, Georgia Institute of Technology, 837 State Street, Atlanta,
  GA 30332; \href{mailto:aycin.aykutalp@gatech.edu}{aycin.aykutalp@physics.gatech.edu}}}

\affil{$^{2}${Scuola Normale Superiore, 
	Piazza dei Cavalieri 7, I-56126 Pisa, Italy}}

\affil{$^{3}${Kapteyn Astronomical Institute,
  University of Groningen, PO Box 800, 9700 AV Groningen, The Netherlands
  }}

\affil{$^{4}${Leiden Observatory, Leiden University,
	P.O. Box 9513, NL-2300 RA  Leiden, The Netherlands}}
	
\begin{abstract} 
In the last decade, the growth of supermassive black holes (SMBHs) has been intricately linked to galaxy formation and evolution and is a key ingredient in the assembly of galaxies. To investigate the origin of SMBHs, we perform cosmological simulations that target the direct collapse black hole (DCBH) seed formation scenario in the presence of two different strong Lyman-Werner (LW) background fields. These simulations include the X-ray irradiation from a central massive black hole (MBH), $\rm{H}_2$ self-shielding and stellar feedback from metal-free and metal-enriched stars. We find in both simulations that local X-ray feedback induces metal-free star formation $\sim 0.5$ Myr after the MBH forms. The MBH accretion rate reaches a maximum of $10^{-3}$ $M_{\odot}$ yr$^{-1}$ in both simulations. However, the duty cycle differs which is derived to be $6\%$ and $50\%$ for high and low LW cases, respectively. The MBH in the high LW case grows only $\sim 6\%$ in 100 Myr compared to $16\%$ in the low LW case. We find that the maximum accretion rate is determined by the local gas thermodynamics whereas the duty cycle is determined by the large scale gas dynamics and gas reservoir. We conclude that radiative feedback from the central MBH plays an important role in star formation in the nuclear regions and stifling initial MBH growth, relative to the typical Eddington rate argument, and that initial MBH growth might be affected by the local LW radiation field.  

\end{abstract}


\section{INTRODUCTION}

Observations of high redshift ($z>6$) quasars suggest that these objects are powered by SMBHs with masses on the order of $10^9$ $M_{\odot}$ \citep{2003AJ....125.1649F, 2006AJ....131.1203F, 2007ApJ...669...32K, 2011Natur.474..616M}. Fundamental to understanding their existence within the first billion years after the Big Bang, is the identification of their formation processes, growth rate and evolution through cosmic time. There are three plausible scenarios for the formation of these seed black holes: a) they are the remnants of the first stars ($M_{\rm{BH}} \sim 10^2$ $M_{\odot}$, \citealt{2003ApJ...582..559V, 2005ApJ...633..624V, 2007MNRAS.374.1557J}), b) they formed in dense star clusters through mergers ($M_{\rm{BH}} \sim 10^4$ $M_{\odot}$, \citealt{1978MNRAS.185..847B, 2001ApJ...562L..19E}), and c) they formed in the isothermal direct gaseous collapse in atomic cooling halos ($M_{\rm{BH}} \sim10^4-10^6$ $M_{\odot}$, \citealt{1993MNRAS.263..168H, 1993ApJ...419..459U,2003ApJ...596...34B,  2006ApJ...652..902S, 2008ApJ...682..745W}) in pre-galactic objects. 

In this work we simulate the latter scenario for the formation of a MBH ($M_{\rm{MBH}} \sim10^4-10^6$ $M_{\odot}$). In order to form a MBH through direct collapse, fragmentation into stars needs to be prevented. Here, it is important that the gas cloud has primordial composition. Metal-enriched gas cools very efficiently by fine-structure metal lines to temperatures of $T<100$ K, causing gas cloud to fragment into smaller clumps of $\leq 100 M_{\odot}$, since the Jeans mass ($M_{\rm{J}}$) scales with the temperature of the ambient gas as $M_{\rm{J}} \propto T^{3/2}$. In metal-free gas the dominant coolant below 10$^4$ K is molecular hydrogen, H$_2$, which can cool the gas down to 200 K through ro-vibrational transitions. Thus, to avoid fragmentation the formation of H$_2$ needs to be suppressed. Indeed, in their studies  \cite{2003ApJ...596...34B} and \cite{2006ApJ...652..902S} have shown that in the absence of H$_2$ fragmentation is sufficiently prevented. 

H$_2$ is fragile and can easily be dissociated by photons in the LW bands (E $= 11.2 - 13.6$ eV). Recent works on the photodissociation of H$_2$ in protogalaxies \citep{2010MNRAS.402.1249S, 2011MNRAS.418..838W, 2014MNRAS.443.1979L,2014arXiv1406.1465L} have shown that a LW flux in the range of $10^2-10^5$ $J_{21}$, where $J_{21}$ is the specific intensity just below 13.6 eV ($J_{21} = 10^{-21}$ erg s$^{-1}$ cm$^{-2}$ sr$^{-1}$ Hz$^{-1}$), is sufficient to prevent H$_2$ formation, and hence fragmentation, in halos with $T_{\rm{vir}} = 10^4-10^5$ K. Moreover, \cite{2012MNRAS.422.2539I} and \cite{2014MNRAS.442L.100V} have argued that, there is a zone of no-return for a collapsing gas cloud to form a DCBH without fragmenting into smaller clumps, which depends on the density ($n>10^4$ cm$^{-3}$) and temperature ($T \geq 10^4$ K ) of the collapsing gas. The latter authors also emphasized that delaying H$_2$ cooling in a collapsing gas clouds, in the absence of a LW background radiation, will not be enough to avoid fragmentation, since throughout the collapse the density of the cloud will increase and H$_2$ cooling eventually take over.

After the seed black hole forms, it mainly grows through gas accretion, where the gas reservoir depends on mergers and/or galaxy interactions. The accretion of gas onto the central black hole yields a luminous radiation source with a broad spectrum. As studied in \cite{2013ApJ...771...50A} (hereafter Paper I), the thermodynamics of the gas in the inner region of an active galactic nucleus (AGN) is dominated by the X-ray radiation produced by the infall of gas onto the central MBH. Thus, to understand the growth of MBH it is crucial to include the X-ray radiation from the AGN. In order to incorporate the effects of X-rays on the MBH growth and on star formation around an AGN, in a self-consistent manner, we have implemented the XDR chemical network of \cite{2005A&A...436..397M} into Enzo (for details, see Paper I).

The aims of this paper are, for the DCBH formation scenario, (1) to investigate the possible connection between SMBHs and the formation and evolution of atomic cooling halos, (2) to follow the accretion history of the seed MBH, and (3) to assess the effects of episodic X-ray irradiation on the ambient gas and hence, on the growth of the seed MBH. This paper is structured as follows. In Section 2, we describe our simulation setup. We present our results and implications in Section 3. Finally, in Section 4, we discuss our findings and possible caveats in our simulations.

\section{SIMULATIONS}

\indent In this work, we use the Eulerian adaptive mesh refinement (AMR) hydrodynamic code Enzo \citep{2014ApJS..211...19B}, version 2.1, which is modified to include metallicity dependent XDR physics and H$_2$ self-shielding. We perform simulations in a three-dimensional periodic box with a side length of 3 h$^{-1}$ Mpc, initialized at $z=99$ by using $\it{inits}$, a package that uses Zel'dovich approximation and is included with the Enzo distribution. The size of the root grid is 128$^3$ with three nested subgrids, each refined by a factor of two. The finest grid has an effective resolution of 1024$^3$ with a side length of 375 h$^{-1}$ kpc. Refinement is restricted to the finest grid and occurs on baryon overdensities of $3\times 2^{-0.2l}$. Here l is the AMR level, and the negative exponent means that the mass resolution in the calculations is super-Lagrangian \citep{2008ApJ...673...14O}. The maximum level of refinement that is reached in the finest grid is 10, allowing us to have a resolution of 3.6 proper pc. The virial mass of the most massive halo at redshift $z= 15$ is $M_{vir} = 2.2 \times 10^8 M_{\odot}$, where $M_{vir}$ is the mass in a sphere that encloses an average dark matter overdensity of 200.

We use Wilkinson Microwave Anisotropy Probe seven-year cosmological parameters (Komatsu et al. 2009): $\Omega _{\Lambda}$ = 0.734, $\Omega _{m} = 0.266$,  $ \Omega _b$ = 0.0449, $\sigma_8$ = 0.81, and $h=0.701$. Here, $\Omega _{\Lambda}$ is the vacuum energy, $\Omega _{m}$ is the matter density, $\Omega _b$ is the baryon density, $\sigma_8$ is the variance of random mass fluctuations in a sphere of radius 8 $h^{-1}$ Mpc, and $h$ is the Hubble parameter in units of 100 km s$^{-1}$ Mpc$^{-1}$.

\subsection{$\rm{H}_2$ Self-Shielding \& Black Hole Accretion}

\indent For a MBH to form at $z=15$ in the direct collapse scenario, its birth cloud cannot fragment through H$_2$ cooling as it collapses. Therefore, we introduce a strong LW background radiation field that could originate from a nearby galaxy \citep{2008MNRAS.391.1961D, 2014MNRAS.442.2036D, 2012MNRAS.425.2854A, 2013MNRAS.432.3438A}. We perform two simulations, one with a LW background intensity of 10$^3\, J_{21}$ (hereafter  $BG_3$) and one for 10$^5\, J_{21}$ (hereafter $BG_5$). We turn on the LW background at $z=30$ and keep it on for the remainder of the simulations. Recent works by \cite{2011MNRAS.418..838W} and \cite{2014arXiv1407.4472R} have shown that such $J_{\rm{LW}}$ values are sufficient to keep the halo almost H$_2$ free.

\indent H$_2$ self-shielding is crucial for the formation of stars in regions where the H$_2$ column densities exceed $10^{14}$ cm$^{-2}$. Thus, we take into account H$_2$ self-shielding as well as the attenuation of the MBH radiation from H \textsc{i}, He \textsc{i}, and He \textsc{ii}.  For H$_2$ self-shielding, we use a local approximation and multiply the H$_2$ photo-dissociation rate by a self-shielding factor $f_{\rm{sh}}$
\begin{eqnarray}
f_{\rm{sh}} &=& \rm{min}\bigg[1,\bigg(\frac {N_{\rm{H_2}}}{10^{14} \rm{cm}^{-2}}\bigg)^{-3/4}\bigg],\\
N_{\rm{H_2}} &=& f_{\rm{H_2}} n_{tot} \lambda_{\rm{J}},
\end{eqnarray}
to correct the impinging UV flux in the LW band \citep{1996ApJ...468..269D, 2010MNRAS.402.1249S}. Here, $N_{\rm{H_2}}$ is the H$_2$ column density, $n_{tot}$ is the total particle number density, and $\lambda_{\rm{J}}$ is the jeans length ($\lambda_{\rm J} = \sqrt{\frac{15 k_b T}{4\pi G\mu\rho}}$). We note that in their work, \cite{2011MNRAS.418..838W} have shown that using the $\lambda_{\rm{J}}$ causes underestimating the $N_{\rm{H2}}$ about an order of magnitude in the low density regimes ($n<10^4$ cm$^{-3}$), and Sobolev length \citep{1957SvA.....1..678S}, which takes into account the velocity gradients in the gas, is a more accurate method (see the discussion section for further details).

\indent At redshift $z=15$ we insert a MBH with a mass of $ M= 5 \times 10^4 M_{\odot}$. We include the radiative feedback from the MBH using the Enzo radiation transport module $\it{Moray}$ \citep{Wise11_Moray}. The radiation from the MBH is modeled as follows. To calculate the gas temperature, we use an XDR grid of models produced for a large parameter space in X-ray flux $F_{\rm{X}}$, number density $n$, column density $N_{\rm{H}}$, and metallicity $Z/Z_{\odot}$ (see section \ref{sec:xdr}). We employ $\it{Moray}$ to compute the full (chemical, thermal and hydrodynamic) response of X-ray exposed gas. We use an $N_{\rm{HI}}$ lookup-table for a polychromatic X-ray spectrum to calculate the attenuation in each line of sight \citep{2006NewA...11..374M}. The radiative transfer equation is numerically solved before the simulation, giving a relative ionizing photon flux I$_\nu$ as a function of the neutral hydrogen column density $N_{\rm{H}}$. The relative ionizing photon flux for H$\textsc i$, He$\textsc i$, and He$\textsc {ii}$ is computed and stored for 300 column densities, equally log-spaced over the range $N_{\rm H}=10^{12}-10^{25}$ cm$^{-2}$. The details of this approach are described in Paper I.

\indent For the accretion of gas onto the MBH, we use the prescription of \cite{2011ApJ...738...54K}. We calculate the accretion rate by using the Eddington-limited spherical Bondi-Hoyle equation
\begin{eqnarray}\label{eq:e3}
\dot{M}_{\rm{BH}} &=& \rm{min} (\dot{M}_{\rm{B}}, \dot{M}_{\rm{Edd}})\\
&=&\rm{min} \bigg(\frac{4\pi G^2 M_{\rm{BH}}^2\rho_{\rm{B}}}{c_s^3}, \frac{4\pi G M_{\rm{BH}}m_{\rm{p}}}{\epsilon \sigma_{\rm{T}}c}\bigg),\nonumber
\end{eqnarray}
where G is the gravitational constant, $M_{\rm{BH}}$ is the mass of the MBH, $\rho_{\rm{B}}$ is the density at the Bondi radius, c$_s$ is the sound speed, $m_{\rm{p}}$ is the mass of a proton,  $\epsilon$ is the radiative efficiency, and $\sigma_{\rm{T}}$ is the Thomson scattering cross-section. And the Bondi radius 
\begin{eqnarray}
R_{\rm{B}} &=& \frac{2 G M_{\rm{BH}}}{c_{\rm{s}}^2}.
\end{eqnarray}
The gas inside a Bondi radius is accreted onto the MBH and it is uniformly subtracted from grid cells. 

\subsection{Star Formation \& Feedback}\label{sec:sf}

In order to model the interplay between stellar and black hole feedback, we employ different recipes for Pop III and PopII/I star formation. We allow star formation to occur only in the finest AMR levels. In our simulation a Pop III star particle, representing a single star, is created when all of the following criteria are met \citep{2007ApJ...659L..87A, 2008ApJ...685...40W, 2012ApJ...745...50W}:
\begin{itemize}
\item[(1)] the gas overdensity of $5 \times10^5$ ($\sim 3 \times10^{3}$ cm$^{-3}$ at $z=15$),
\item[(2)] the metallicity of the gas $<10^{-3.5}$ solar,
\item[(3)] the molecular hydrogen fraction $\rm{f}_{\rm{H_2}}  < 5 \times 10^{-4}$,
\item[(4)] the cooling time is less than the dynamical time, $t_{\rm cool}$ $<$ $t_{\rm dyn}$, 
\item[(5)] the velocity flow is converging; i.e., $\nabla \cdot \bold v < 0$.
\end{itemize} 

For the Pop II/I star formation recipe we use the algorithm of \cite{2012ApJ...745...50W} in Enzo. In the simulations, a Pop II/I star particle, representing a stellar cluster,  is formed when the following criteria are met: (1) the gas overdensity  $5 \times10^5$, (2) the metallicity of the gas $Z >10^{-3.5} Z_{\odot}$, (3) $t_{\rm cool}$ $< t_{\rm dyn}$, (4) the velocity flow is converging; i.e., $\nabla \cdot \bold v < 0$. In our simulations, the minimum mass of a Pop II stellar cluster is $10^3 M_{\odot}$. For every star particle that is created, ionizing radiation transport is included following the prescriptions of \cite{2012ApJ...745...50W}, i.e., every star particle radiates UV photons for 20 Myr. The energy injection by the supernova explosions (SNe) of Pop III stars is computed from the stellar mass and is deposited in a sphere of 10 pc radius. We follow the metal production by SNe and take into account the metal cooling. 

\subsection{XDR Grids}\label{sec:xdr}

We tabulate the XDR grids using a modified version of the chemical network of \cite{2005A&A...436..397M}. Our chemical network consists of 176 species and more than 1000 reactions.

The main heating mechanism in XDRs is Coulomb heating when fast electrons interact with thermal electrons, where the heating efficiency is as high $10-50\% $ \citep{1996ApJ...466..561M}. Moreover, X-rays have small absorption cross-sections, which roughly scale as $1/E^3$, and thus they can penetrate large columns. An important quantity in XDRs that affects the chemo-thermal state of the gas is the energy deposition rate in the gas parcel. In Paper I, we found that the energy deposition rate in a solar metallicity gas is much higher than for the zero metallicity case, causing high temperatures ($\sim 10^6$ K) in the central 40 pc. Furthermore, earlier work on the X-ray effects from an AGN by \cite{2011ApJ...730...48P} has shown that X-ray exposed molecular gas has temperatures five times higher than gas in a starburst of equal bolometric power. This has important consequences for the initial mass function (IMF) of stars, because the Jeans Mass ($M_{\rm{J}}$) scales with the temperature of the ambient gas as $M_{\rm{J}} \propto T ^{3/2}$.

In XDRs at high temperatures (T $>$ 5000 K), the gas cooling is dominated by collisional excitation of Ly$\alpha$, and forbidden transitions of [O I] ($\lambda \lambda$ 6300, 6363 $\mu$m), [C I] ($\lambda \lambda$ 9823, 9850 $\mu$m), [Fe II]($\lambda \lambda$ 1.26, 1.64 $\mu$m), and [Si II] ($\lambda \lambda$ 6716, 6731$\mu$m). At low temperatures ($T<3000$ K), gas cooling is dominated by the fine-structure lines of [OI] 63 $\mu$m, [SiII] 35 $\mu$m, [CII] 158 $\mu$m, [CI] 269 and 609 $\mu$m, as well as rotational lines of CO and H$_2$O. For further details, we refer the interested reader to \citet{2005A&A...436..397M} and Paper I.

We construct and utilize tables for species abundances and gas temperatures over a wide range of X-ray flux $F_{\rm{X}} = 10^{-1.25}-10^{5.5}$ erg cm$^{-2}$ $\rm{s}^{-1}$, number density $n = 10-10^6$ cm$^{-3}$, column density  $N_{\rm{H}} = 10^{20}-10^{24}$ cm$^{-2}$, equally spaced with a step-size of 0.25 dex, and metallicity Z/Z$_{\odot} =10^{-6},10^{-4}, 10^{-2},1$. This large parameter space enables us to model the interstellar medium (ISM) properties close to an AGN properly. We use Enzo's 9 species (H, H$^+$, H$^-$, He, He$^+$, He$^{2+}$, H$_2$, H$_2^-$, and e$^-$) non-equilibrium chemical network for zero metallicity gas \citep{1997NewA....2..181A,1997NewA....2..209A} in regions that are not X-ray dominated. 

\section{RESULTS \& IMPLICATIONS}

In order to investigate the effects of X-ray irradiation on the growth of MBHs and on the formation and evolution of stellar population in the host halo we perform cosmological, radiation hydrodynamics simulations. The growth of MBHs is simulated under the influence of two different LW background radiation fields: 10$^3$ and 10$^5$ $J_{\rm{21}}$. Here, we assume that there is a close by star-forming halo (Pop II stars with average surface temperatures $T=10^4$ K) within 10 kpc which could provide such high LW fluxes \citep{2008MNRAS.391.1961D}. We insert the seed MBH with a mass of $M_{\rm{BH}} = 5 \times 10^4 M_{\odot}$ into the center of our favorite halo ($M_{\rm{H}} = 2 \times 10^8 M_{\odot}$) at $z= 15$. However, we turn on the radiation field, H$_2$ self-shielding and the star formation modules at $z= 30$, before we insert our MBH to the center of our halo. This is done to check whether the conditions for the DCBH formation scenario are met in our favorite, hosting dark matter halo. We only follow growth of DCBHs for 100 Myr, due to high computational expenses of the radiative transfer.

\subsection{High LW Case ($BG_5 = 10^5 J_{\rm{21}}$)}
 
\begin{figure*}[!htb]
\centering
\includegraphics[angle=0,width=12cm]{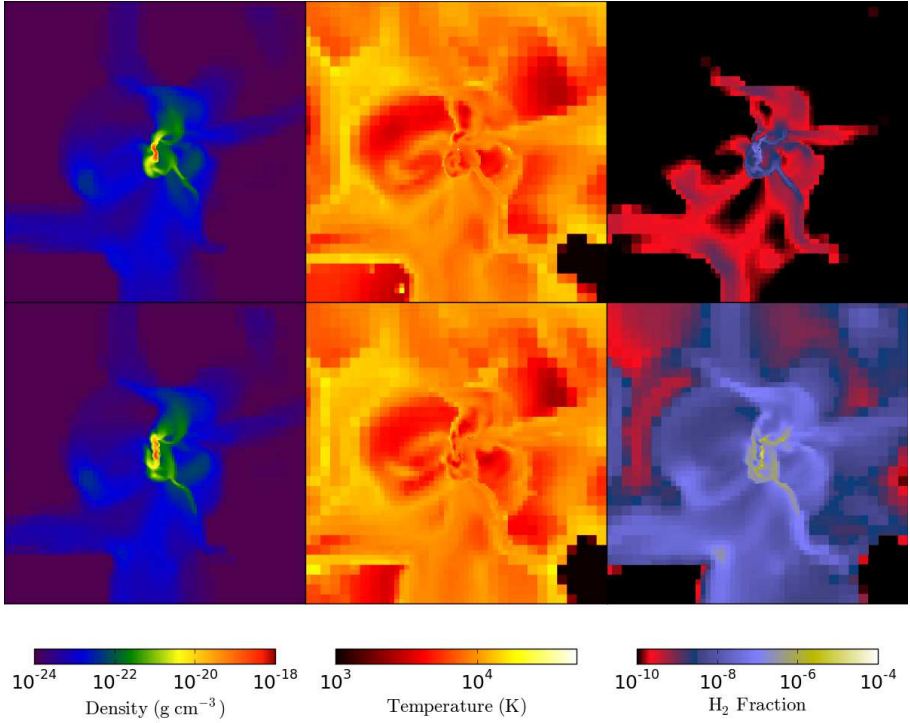}
\caption {Gas density, temperature and H$_2$ fraction slices in the y-plane at $z=15$ for the $BG_5$ (top) and $BG_3$ (bottom) runs with a 1 kpc field of view. Prior to the DCBH formation the only difference in between the two simulations is the H$_2$ fraction of the hosting halo.}\label{fig:Fig1}
\end{figure*}

The only initial difference between the $BG_5$ and $BG_3$ runs is the strength of the LW background radiation that photo-dissociates H$_2$ very efficiently. Thus, prior to the DCBH formation at $z=15$,  the H$_2$ fraction ($f_{\rm{H_2}}$) in the inner 100 pc of our host halo is as low as $10^{-8}$. This is shown in Figure \ref{fig:Fig1}, where we plot gas density, temperature and H$_2$ fraction slices in the y-plane through the densest point at $z=15$ for the $BG_5$ (top) and $BG_3$ (bottom) runs. The corresponding H$_2$ column density (for a typical density of 10$^{-21}$ g cm$^{-3}$) over a scale of 30 pc is $10^{15}$ cm$^{-2}$ ($f_{sh}=0.17$). Thus, self-shielding occurs but it is not very strong yet, given the broad line wings in the H$_2$ dissociative bands. Under these initial conditions, in our atomic cooling halo with $T_{\rm vir} \sim 10^4$ K, we therefore do not form any Pop III stars prior to $z = 15$. Hence, our DCBH scenario for the formation of the seed MBH is appropriate. 

\begin{figure}
\includegraphics[angle=0,width=9cm]{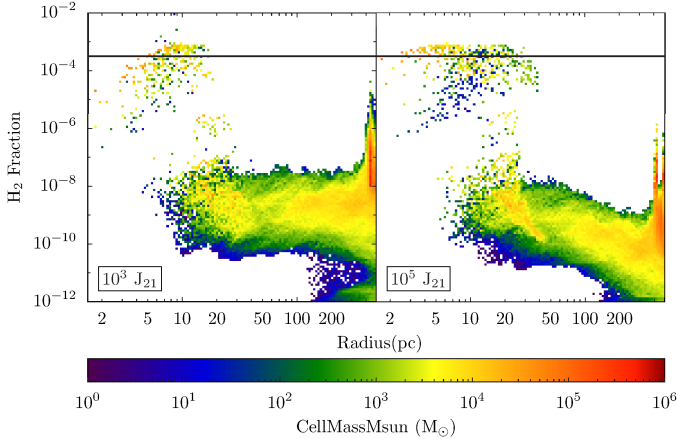}
\caption {H$_2$ fraction vs. radius for $BG_3$ (left) and $BG_5$ (right) cases, 1.265 Myrs after the seed MBH is inserted. The black horizontal line correspond to our Pop III star formation criteria. There is more gas at $\rm{r} >10$ pc which is above the Pop III star formation threshold in the $BG_3$ case. This is due to the fact that X-rays penetrate further, $\sim 500$ pc, in the $BG_5$ case.}\label{fig:Fig2}
\end{figure}

In the center of our halo, the MBH immediately starts accreting and emitting X-ray radiation. This X-ray radiation increases $e^-$ fraction which enhances H$_2$ formation in zero-metallicity gas through the H$^-$ route
\begin{eqnarray}
H + e^- \rightarrow H^- + \gamma, \\
H + H^- \rightarrow H_2 + e^- .
\end{eqnarray}
In Figure \ref{fig:Fig2}, we plot the H$_2$ fraction in the inner 500 pc for both $BG_5$ (right) and $BG_3$ (left) cases, $\sim1.3$ Myr after the seed MBH was inserted. After being irradiated for $\sim 0.5$ Myr, the H$_2$ abundance increases to $5 \times 10^{-4}$, which is the criterion for Pop III star formation in our simulations, and this nuclear region experiences X-ray induced Pop III star formation. Note that there is gas at $r \sim 20-40$ pc with higher $f_{\rm{H_2}}$ than our Pop III criterion, which is absent in the $BG_3$. This is due to the fact that X-rays penetrate further, $\sim 500$ pc, in the $BG_5$ case and thus boost the $f_{\rm{H_2}}$. In order to check whether this rapid Pop III formation is purely due to the X-ray irradiation from the central MBH we conducted another simulation with the same setup but without X-ray radiation. In this simulation Pop III stars formed only $\sim 10$ Myrs after we insert the MBH. Hence, we conclude that X-ray irradiation from the central MBH initially has a positive feedback effect on the star formation, by inducing H$_2$ formation through the H$^-$ route, and thus it is very important to take into account not only for the MBH growth but also for regulating the stellar population in the host halo.

\begin{figure*}
\centering
\includegraphics[angle=0,width=12cm]{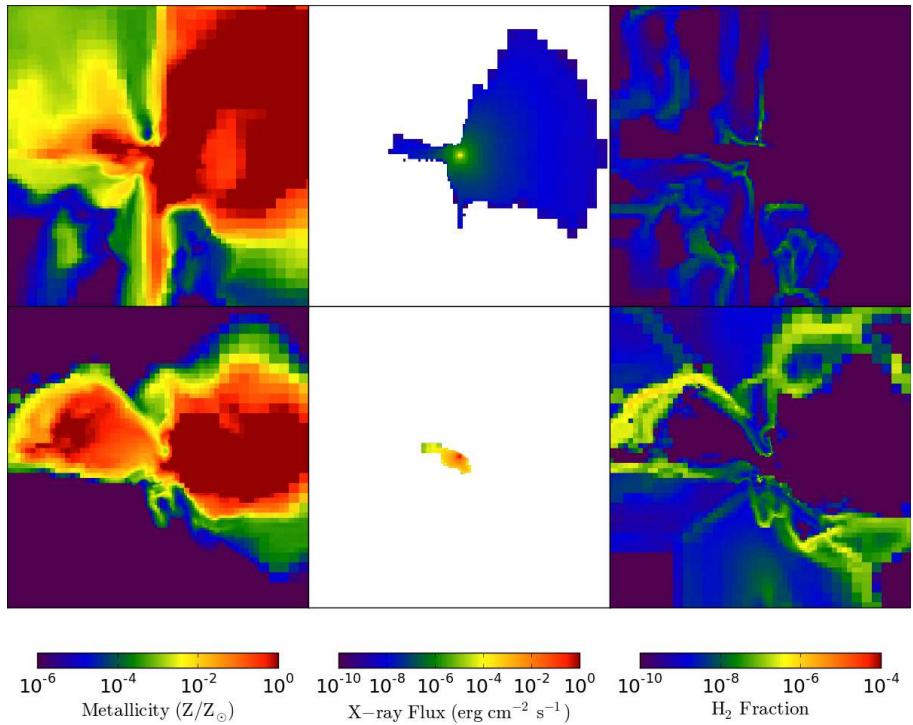}
\caption {Metallicity, X-ray flux and H$_2$ fraction slices in the y-plane with a 1 kpc field of view through the centre of the MBH for $BG_5$ (top) and $BG_3$ (bottom) runs at $z=14.86$ when Pop III SNe first occur in the halo(3.6 Myrs after the MBH is inserted). In the $BG_5$ case X-rays penetrate further, $\sim 500$ pc.}\label{fig:Fig3}
\end{figure*}

In Figure \ref{fig:Fig3}, we plot metallicity, X-ray flux and H$_2$ fraction slices of 1 kpc through the centre of the MBH for $BG_5$ (top) and $BG_3$ (bottom) cases at $z=14.86$ (3.6 Myr after the MBH is inserted), when Pop III SNe first occur in the halo. Due to the strong LW background, the H$_2$ abundance and thus cooling in the halo is efficiently suppressed, leading to high temperatures and low densities. Thus, the strong LW background renders the host halo fragile to radiative feedback both from SNe and the MBH and thus the collective effect on the ambient gas is hostile. As illustrated in Figure \ref{fig:Fig3}, the SNe blast-waves heat and destroy any H$_2$ in their wake and subsequently chemically enriches the surrounding medium. X-ray radiation penetrates up to 500 pc from the center, due to the very strong LW background leading low column densities and optical depth for X-rays. When the SNe go off, they form H\textsc{ii} regions which propagate much faster in the BG5 case. This is because for a local ionizing photon rate $S_i$ , the Str\"{o}mgren radius scales as $R_{S}\propto (S_i/n^2 )^{1/3}$ , the recombination time scales as $t_s \propto1/n_e$, and the ionization front velocity scales as $V_I\propto S_i/n \times R^2$. This in turn, provides a path for metals to enrich the ISM up to $\sim1$ kpc.

\begin{figure*}[!htb]
\includegraphics[angle=0,width=8.2cm]{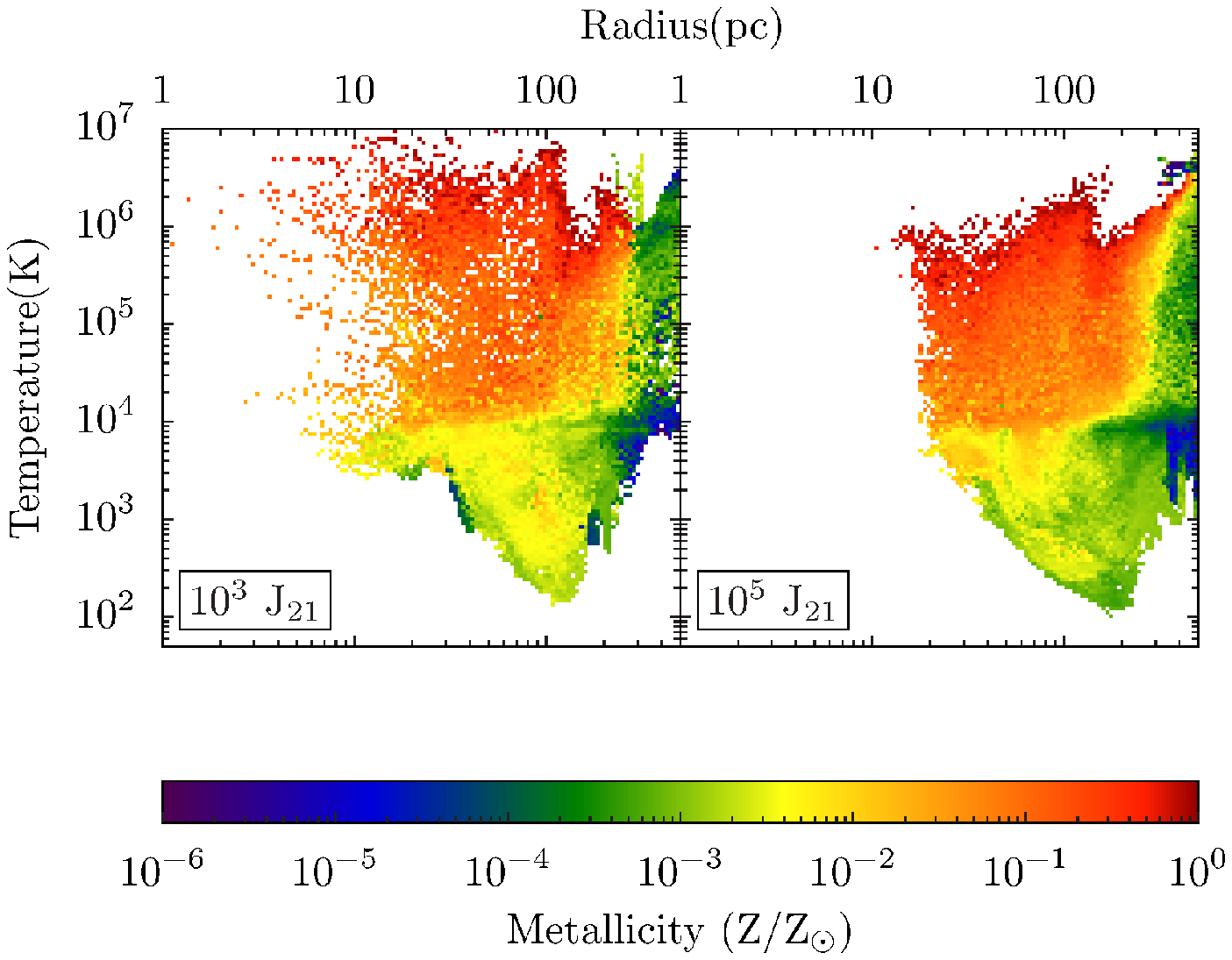}
\vspace{0.01 cm}
\includegraphics[angle=0,width=9.0cm]{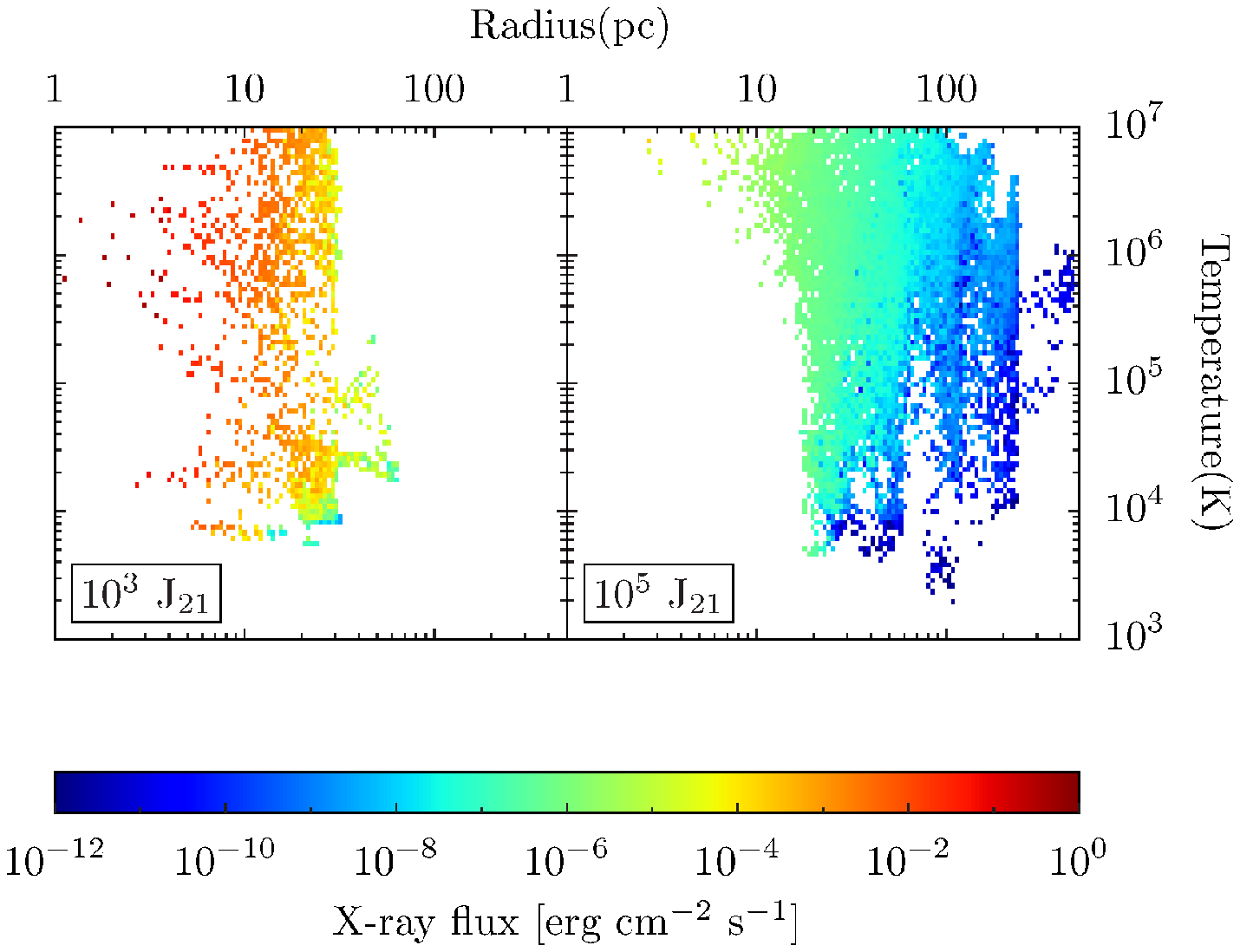}
\\
\center
\includegraphics[angle=0,width=9.5cm]{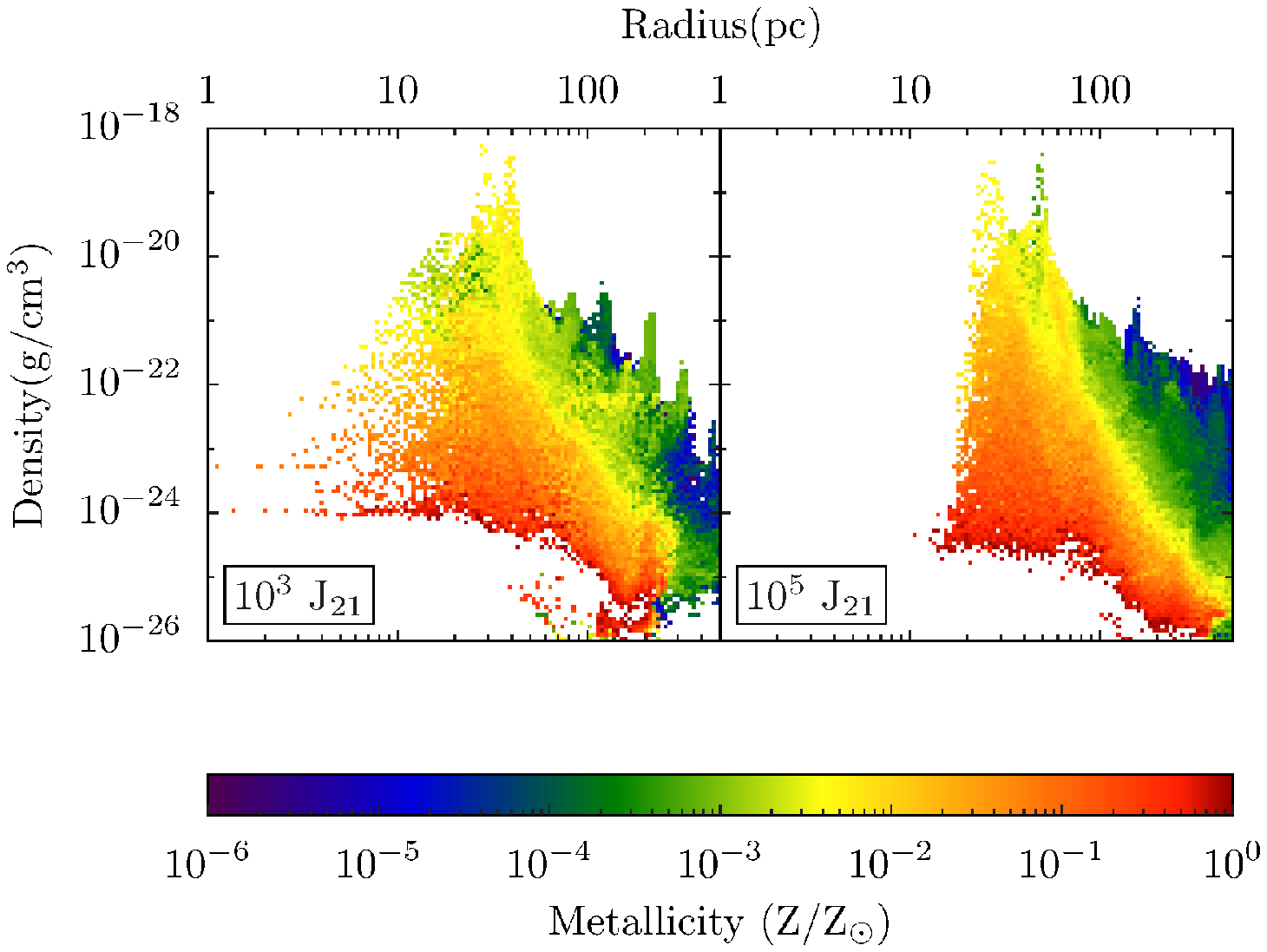}
\caption {Top panel: temperature versus radius color-coded by metallicity (Ieft panel) and X-ray flux (right panel) for both $BG_5$ and $BG_3$ simulations at $z=14.86$ (3.6 Myr after the MBH is inserted). The SNe blast-waves propagate faster in the $BG_5$ case and blow away the gas from the inner $\sim 15$ pc. Bottom panel: density versus radius color-coded by metallicity. The temperature of the high density high-metallicity gas is higher than the low-metallicity gas due to the large cross-sections available to X-rays penetrating metal-rich gas.}\label{fig:Fig4}
\end{figure*}

In Figure \ref{fig:Fig4}, we plot temperature versus radius color-coded by metallicity (left panel) and X-ray flux (right panel) for both simulations. The SNe blast-waves propagate faster in the $BG_5$ case and blow away the gas from the inner $\sim 15$ pc. Although the X-ray irradiation from the MBH is weak, due to the lack of gas, it penetrates to larger distances from the MBH with respect to the $BG_3$ case (right panel of Figure \ref{fig:Fig4}).

\begin{figure*}[!htb]
\includegraphics[angle=0,width=8.8cm]{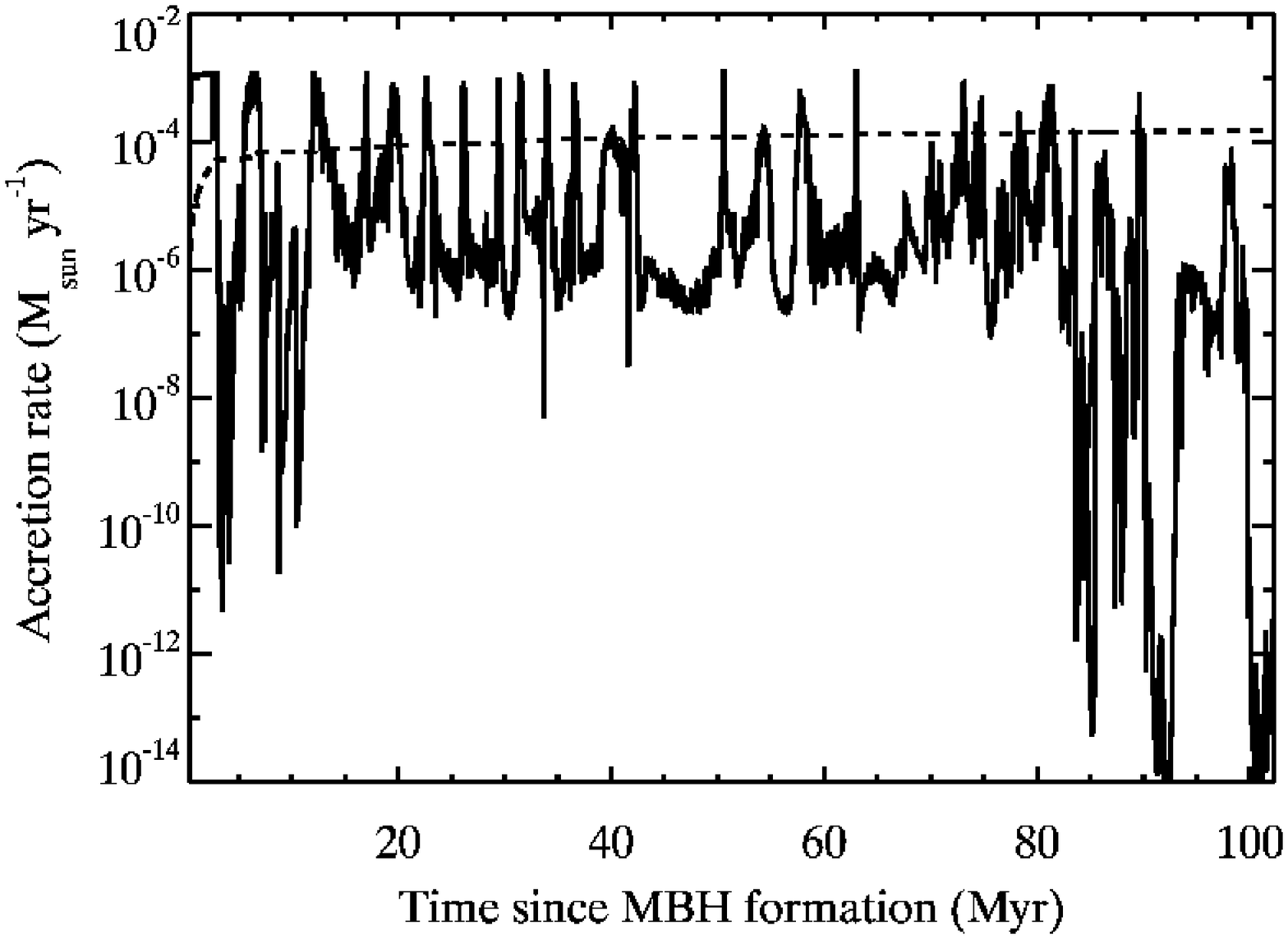}
\vspace{0.01 cm}
\includegraphics[angle=0,width=8.7cm]{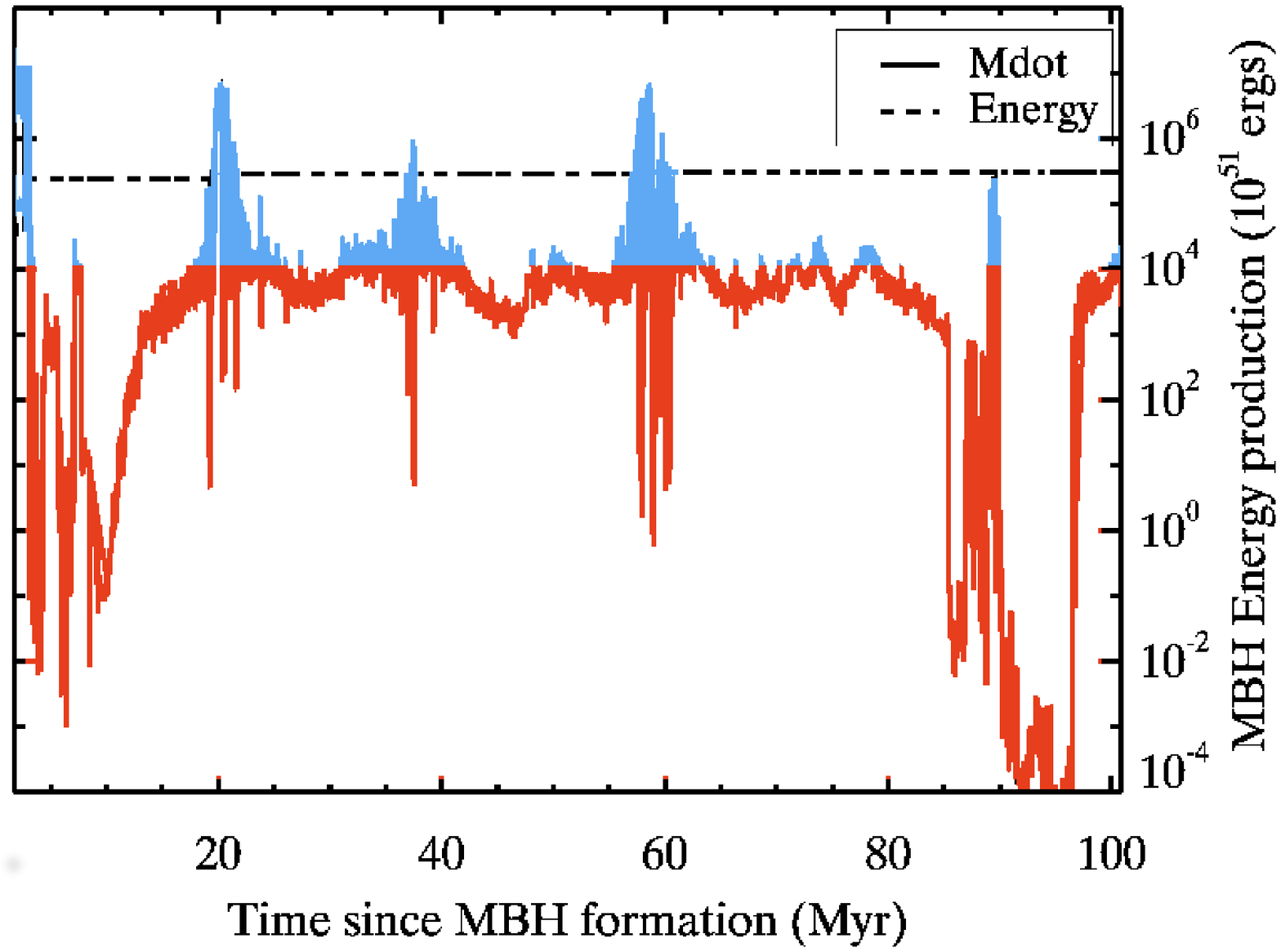}
\vspace{0.01cm}
\caption {Accretion rate and the corresponding total energy production of the central MBH over time for the $BG_3$ (left) and $BG_5$ (right) runs. The maximum accretion rate that both MBHs experience is similar, $10^{-3}\rm{M}_{\odot}$ yr$^{-1}$. However, the duty cycle of these two MBHs is derived to be $6\%$ and $50\%$ for the $BG_5$ and $BG_3$ cases, respectively. The red-colored part on the $BG_5$ plot shows the hot gas ($T\geq10^4$ K) dominated accretion whereas the blue-colored part shows the cold gas ($T<10^4$ K) dominated accretion.}\label{fig:Fig5}
\end{figure*}

Figure \ref{fig:Fig5} shows the accretion rate (solid) and corresponding total energy production (dashed) for both $BG_3$ (left) and $BG_5$ (right) runs for 100 Myr. The MBHs initially accrete gas efficiently at a rate $\sim 10^{-3}$ $M_{\odot}$ yr$^{-1}$ ($10\%$ Edddington rate), 2 Myr after they are inserted into the simulation.
This accretion rate corresponds to a total energy production of $2 \times 10^{56}$ erg over 1 Myr at radiative efficiency $\epsilon = 0.1$, which is about the same as the energy produced by $2\times10^5$ SNe explosions. The red-colored part in the right-panel of Figure \ref{fig:Fig5} shows the hot gas ($T\geq10^4$ K) dominated accretion rate whereas the blue-colored part shows the cold gas $T<10^4$ K) dominated accretion rate. This implies that sound speed ($c_{s}^3$) in the equation \ref{eq:e3}  is more strongly dependent on the temperature of the gas ($ \propto T^{-3/2}$) than the density of the gas  ($\rho_B$). Overall gas densities in the inner 30 pc are 2-3 orders of magnitude lower ($\rho \sim 10^{-23 }$ g cm$^ {-3}$) than the $BG_3$ case. Thus, the response time of the ambient gas to both UV and X-ray radiation is much longer. This also helps to explain the longer duty cycles seen in $BG_5$ run and we discuss this further in Section \ref{sec:LW}.

In order to investigate the dependence of the MBH growth on the local versus large scale gas thermodynamics we perform a simulation where we insert the MBH 10 pc off the center of the potential well of the host halo. In this case, because the MBH does not reside in the most dense region in the halo, the enhancement in the H$_2$ fraction by the X-ray irradiation was not enough to instigate an H$_2$-driven collapse. This is very important because without star formation there is no metal enrichment. 

\begin{figure}
\includegraphics[angle=0,width=8.8cm]{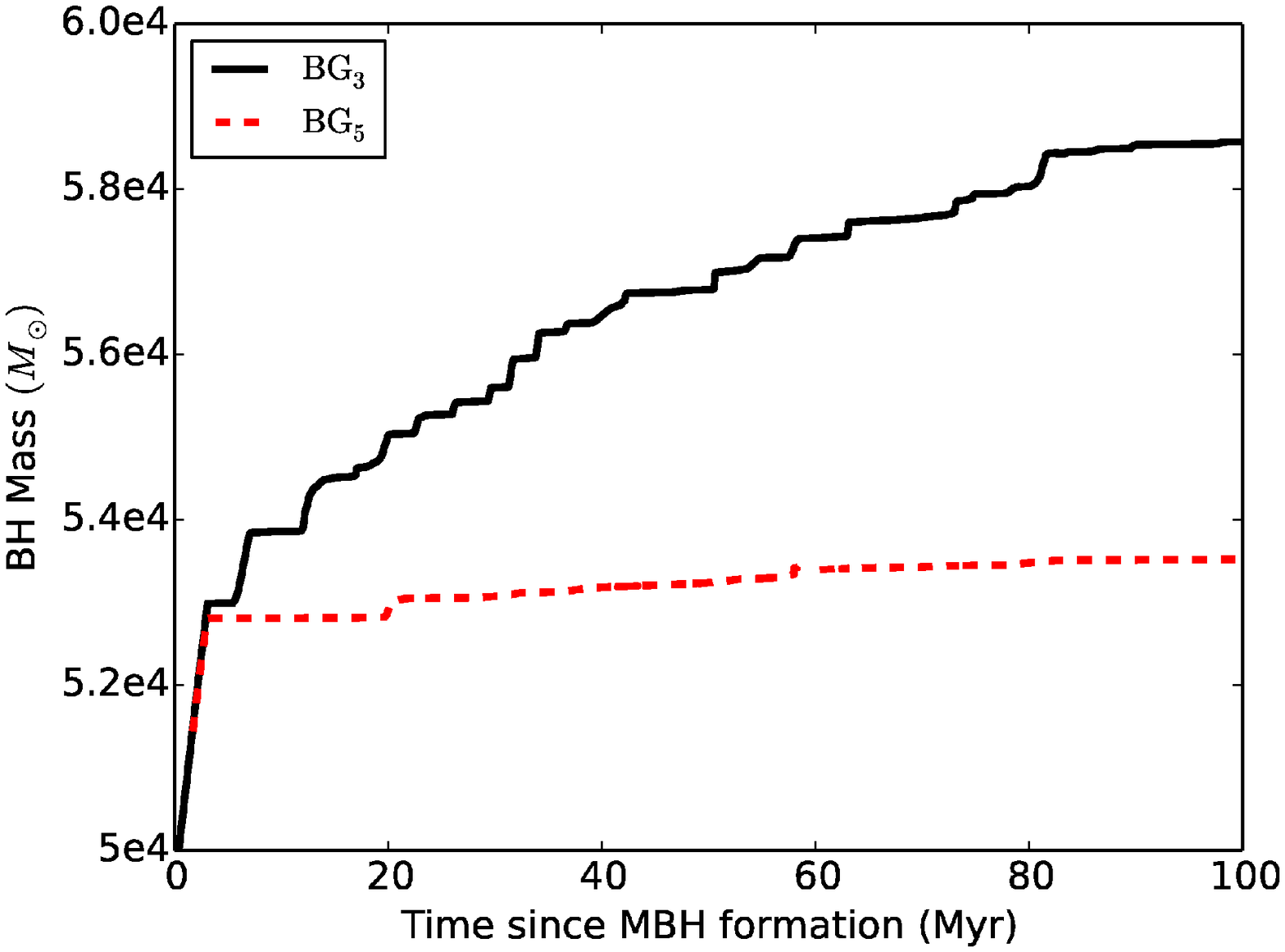}
\caption {Growth of the MBH over 100 Myr for the $BG_5$ (red-dashed line) and $BG_3$ (black-solid line) cases. In either cases, the MBH growth is stifled relative to the typical Eddington rate argument.}\label{fig:Fig6}
\end{figure}

Indeed, in this simulation setup, the enhancement of H$_2$ from the X-ray radiation (through H$^-$ route) was not high enough to induce a collapse, keeping the halo star-free and thus metal-free for 100 Myr. The response of metal-free gas to the high LW radiation, which dissociates H$_2$ and thus keep the gas hot and diffuse, was so long that the accretion rate was 4 orders of magnitude less than the original $BG_5$ case. Thus, the MBH grew only $1\%$ in 100 Myr whereas in the original run it grew $\sim 6\%$ as shown in Figure \ref{fig:Fig6}. 

On the other hand, the duty cycle of the two runs are not so different. We derived the duty cycle of these to simulations to be $4\%$ and $6\%$ for off-center and original run, respectively. Therefore, we conclude that the maximum accretion rate is determined by the local gas thermodynamics whereas the duty cycle is determined by the large scale gas dynamics and gas reservoir of the host halo. However, even in the original $BG_5$ case, the MBH grows only at $10\%$ Eddington rate. This is shown in Figure \ref{fig:Fig6}, where we plot the growth of MBH over 100 Myr for $BG_5$ (red-dashed line) and $BG_3$ (black-solid line) cases.

\subsection{Low LW case ($BG_3 = 10^3 J_{\rm{21}}$)}\label{sec:LW}

In the $BG_3$ case, prior to $z = 15$, the H$_2$ abundance fraction is $\sim 10^{-6}$ (100 times higher than the $BG_5$ case), as shown in Figure \ref{fig:Fig1}. Recent works by \cite{2012MNRAS.422.2539I} and \cite{2014MNRAS.442L.100V}, have shown that in order to form a DCBH a zone of no-return should be reached before runaway cooling begins. This requires not only low $f_{\rm{H_2}}$ ($\leq10^{-6}$), but also high initial densities ($n >10^4$ cm$^{-3}$ ) and temperatures ($T \geq 10^4$ K). Since, the $BG_3$ simulation prior to $z=15$ meets these criteria ($n \sim 10^6$ cm$^{-3}$ and $T>10^4$ K, see Figure \ref{fig:Fig1}), our DCBH scenario still holds. 

After 0.5 Myr, the X-ray radiation from the MBH enhances the H$_2$ abundance to $> 5\times 10^{-4}$, instigating a collapse and fragmentation forming Pop III stars. Metal enrichment by Pop III stars provides a high opacity for X-rays because absorptions by inner shell electrons of C, N and O have large cross sections above 1 keV. In Figure \ref{fig:Fig4} (right panel), the X-ray flux is strongly attenuated by the ISM in the inner 20 pc. X-rays do not penetrate above 30 pc due to the high column densities, on the order of $10^{23-24}$ cm$^{-2}$, that are reached in the center. Therefore, the energy deposition rate into the ISM is higher than in the $BG_5$ case (also see Paper I). Furthermore, as studied in Paper I, due to the large cross-sections available to X-rays penetrating metal-rich gas, the temperature of the high density high-metallicity gas is higher than the low-metallicity gas, also shown in Figure \ref{fig:Fig4}. This is converse of the normal behavior of metals in an X-ray free environment. Under normal conditions, where there are no X-rays, metals are responsible for cooling of the gas. However, when there are X-rays, the metal-rich gas is heated due to the high cross-sections. Therefore, it is very important to take into account X-ray feedback effects when studying BH growth where the ISM is enriched by the metals.

The accretion rate after 1 Myr is 10$^{-3}$ $M_{\odot}$ yr$^{-1}$ (see Figure \ref{fig:Fig5}), similar to the $BG_5$ case. Afterwards the MBH accretes at this high rate more frequently than the $BG_5$ case because the nuclear gas can cool more efficiently through H$_2$ due to the lower LW background. In fact, for the $BG_3$ run we obtain a duty cycle of close to 50\%. The $BG_5$ case yields an order of magnitude less (6\%). These numbers are similar to \cite{2012ApJ...747....9P}. In their terminology, the $BG_5$ and $BG_3$ runs experience ``mode-I'' ($6\%$) and ``mode-II'' ($50\%$) accretion. On the other hand, the MBH in the $BG_3$ case does not double its mass in an Eddington time (45 Myr) either. As shown in Figure \ref{fig:Fig6}, it grows only by $16\%$ in 100 Myr. We conclude that, once radiative feedback from the accreting MBH is taken into account, under the influence of an external LW background, the MBH growth is stifled relative to the typical Eddington rate argument. 

\section{DISCUSSION \& CONCLUSIONS}

In this work, we investigate the role of X-ray radiation on regulating MBH growth and facilitating stellar populations in the host dark matter halo, under the influence of a LW background. Here we summarize the interplay between the physics we included in the simulations. The LW background dissociates and suppresses H$_2$ formation. Once the seed DCBH forms, it starts accreting and producing X-rays. This radiation increases the $e^-$ abundance which induces H$_2$ formation in the primordial gas. Then Pop III stars form in our simulations when cold, H$_2$ rich, and dense clouds form. Their SNe enrich the ISM, allowing gas to cool as usual in the absence of strong X-ray radiation. When the seed DCBH accretes gas it produces X-rays that are attenuated by metal-rich gas causing efficient heating. There is thus an interplay between X-rays and the LW background depending on ambient metallicity and the resulting duty cycle of BH accretion. The highlights of our findings are as follows.

\begin{itemize}
\item The presence of a strong LW background renders a primordial atomic cooling halo of $\sim 10^8 M_{\odot}$ fragile to radiative feedback by SNe and a MBH. 
\item The X-ray irradiation from the central MBH induces the initial star formation and Pop III stars form 0.5 Myr after the seed MBH was inserted. However, in the long term it prevents the MBH from accreting at high rates continuously. 
\item The X-ray feedback and MBH growth is self-regulating. The maximum accretion rate that both MBHs experience is similar, $10^{-3} M_{\odot}$ yr$^{-1}$. On the other hand, the duty cycle of these two MBHs is derived to be $6\%$ and 50$\%$ for high and low LW cases, respectively. The MBH in the high LW case grows only $\sim 6\%$ in 100 Myr whereas in the low LW case the MBH grows $16\%$ in 100 Myr. We find that this is due to the fact that the maximum of the accretion rate is determined by the local gas thermodynamics whereas the duty cycle is determined by the large scale gas dynamics and gas reservoir of the host halo. 
\item The initial chemical enrichment is very crucial for MBH growth. Metal rich gas has shorter response times to X-ray radiative feedback which enables gas to accrete onto the MBH. 
\item Once X-ray feedback effect from the accreting MBHs is taken into account, under the influence of external LW background radiation, the MBH growth is stifled relative to the typical Eddington rate argument. Even if the DCBH starts accreting at the Eddington rate after a few Eddington times and can maintain this accretion through cosmic time, then the resulting mass gain would still be a factor of $\sim e^2$ less. 
\end{itemize}

We here stress that, $N_{\rm{H_2}}$ is a non-local quantity, but because of the high computational expenses of determining self-shielding accurately, we have estimated the H$_2$ self-shielding effect by using a local approximation. This local approximation method has been shown to be accurate within an order of magnitude only \citep{2010MNRAS.402.1249S, 2011MNRAS.418..838W}. \cite{2011MNRAS.418..838W} have shown that the Sobolev length method is more accurate than the Jeans length to estimate the $N_{\rm{H_2}}$, especially for densities $n<10^4$ cm$^{-3}$, which reduces the critical specific intensity needed to keep halos H$_2$ free by a factor of six. In our simulations, gas reaches densities of $n=10^6$ cm$^{-3}$ within the inner 40 pc, and we considered $J_{\rm{21}}$ values that are above the critical value, above which the halo remain H$_{\rm{2}}$-free. Hence, in the scenarios presented here, considering the Sobolev length would not change our conclusions on the formation of MBH. However, it might affect the stellar population in the host halo. 

Our simulations assume a constant LW background, which might cause the lower accretion rates since the source of the LW radiation field and its effect on our halo depends on the position and evolution of the source \citep{2014arXiv1407.4472R}. 

The spatial resolution of our simulations is 3.6 pc hence, everything within $\sim 4$ pc is accreted by the MBH. This might lead us to overestimate the accretion rates. However, the accretion prescription we use in this work, Eddington limited spherical Bondi-Hoyle, does not include the angular momentum of the gas which is crucial in order for gas to fall onto the MBH from the accretion disk. In fact, \cite{2011MNRAS.412..269P} showed that taking into account the angular momentum of the accreting gas can influence the response of the ambient gas to the feedback effect from the MBH. Therefore, in the near future, we are planning to perform higher-resolution simulations with a better accretion prescription that will take into account the angular momentum of the infalling gas with an evolving LW background. 

\acknowledgments
We would like to thank the referee for their comments which made the overall paper clearer. This research was supported by National Science Foundation (NSF) grants AST-1211626  and AST-1333360. AA is grateful to Andrea Ferrara for stimulating discussions on black hole growth. Computations and associated analysis described in this work were performed using the publicly-available \enzo{} code (\url{http://enzo-project.org}) and the \yt{} toolkit \citep[\url{http://yt-project.org};][]{2011ApJS..192....9T}.

\bibliographystyle{apj}   
\bibliography{references}

\end{document}